\documentclass[reprint, prl, amsmath, amssymb, aps, superscriptaddress, showpacs, longbibliography]{revtex4-1}

\usepackage{graphicx, color} 
\usepackage[normalem]{ulem}          

\usepackage[colorlinks=true]{hyperref}
\hypersetup{pdfpagemode=None, linkcolor=black, citecolor=blue, urlcolor=blue}  

\usepackage[T1]{fontenc}
\usepackage[utf8]{inputenc}

\usepackage{amssymb}
\usepackage{amsmath}
\usepackage{commath}
\usepackage{graphicx,bm}
\usepackage{verbatim}
\usepackage{ulem}

\usepackage[textsize=scriptsize, colorinlistoftodos]{todonotes}
\usepackage{xcolor} 

\definecolor{britishracinggreen}{rgb}{0.0, 0.26, 0.15}
\definecolor{bulgarianrose}{rgb}{0.28, 0.02, 0.03}
\definecolor{darkred}{rgb}{0.90,0,0}
\definecolor{darkgreen}{rgb}{0,0.45,.15}
\definecolor{darkblue}{rgb}{0,0,1}
\definecolor{orange}{cmyk}{0,0.6,0.8,0}
\definecolor{lightblue}{rgb}{0.3,0.5,1}
\definecolor{lightgreen}{rgb}{0.4,0.80,.4}

\newcommand{\eF}{\varepsilon_{F}}
\newcommand{\vpb}{v_{\textrm{pb}}}

\newcommand{\Eflow}{E_\mathrm{flow}}
\newcommand{\Econd}{E_\mathrm{cond}}
\newcommand{\wpb}{w_\mathrm{pb}}

\begin{document}

\title{Impurity-controlled vortex mobility and pair-breaking in fermionic superfluid rings}

\author{Bu\u{g}ra T\"uzemen}\email{btuzemen@ifpan.edu.pl}
\affiliation{Institute of Physics, Polish Academy of Sciences, Al. Lotnik\'{o}w 32/46, 02-668 Warsaw, Poland}

\author{Andrea Barresi}\email{andrea.barresi
.dokt@pw.edu.pl}
\affiliation{Faculty of Physics, Warsaw University of Technology, Ulica Koszykowa 75, 00-662 Warsaw, Poland}

\author{Gabriel Wlaz\l{}owski}\email{gabriel.wlazlowski@pw.edu.pl}
\affiliation{Faculty of Physics, Warsaw University of Technology, Ulica Koszykowa 75, 00-662 Warsaw, Poland}
\affiliation{Department of Physics, University of Washington, Seattle, Washington 98195--1560, USA}

\author{Piotr Magierski}\email{piotr.magierski@pw.edu.pl}
\affiliation{Faculty of Physics, Warsaw University of Technology, Ulica Koszykowa 75, 00-662 Warsaw, Poland}
\affiliation{Department of Physics, University of Washington, Seattle, Washington 98195--1560, USA}

\author{Klejdja Xhani}\email{klejdja.xhani@polito.it}
\affiliation{Department of Applied Science and Technology, DISAT, Politecnico di Torino, Italy}

\date{\today}

\begin{abstract}
    
Using time-dependent density functional theory, we study how density and size of impurities govern
dissipation of persistent currents of fermionic superfluid rings in the BCS regime. 
The critical winding number for vortex emission increases with impurity density, but this enhancement is impurity size-dependent and capped by the pair-breaking threshold. Below this vortex-emission threshold, the winding number remains constant while flow energy dissipates through impurity-enhanced pair-breaking. 
Above the threshold, vortex-impurity interactions produce distinct mobility regimes—deflected trajectories, individual pinning, collective pinning, and inter-site hopping, controlled by the impurity size and density, which determine the dominant dissipation channel.
These findings provide design principles for ultracold-atom experiments and insights into vortex-pinning dynamics in neutron-star crusts and superconductors.

\end{abstract}

\maketitle
\label{section:intro}
\section{Introduction}
The flow of a Fermi superfluid through a spatially inhomogeneous medium plays a central role in many physical systems and poses several unresolved theoretical challenges. In such environments, the superfluid fraction is generally reduced below unity, implying that only part of the fluid participates in coherent, collective motion~\cite{PhysRevLett.25.1543,PhysRevA.109.023301,PhysRevLett.36.1151,PhysRevB.76.052503,PhysRevB.83.094530,PhysRevA.111.L011302}. The remaining fraction, effectively decoupled from the superflow, is expected to behave as a normal component even at zero temperature, although it may not contribute to entropy production.
The conditions under which this decoupling occurs 
depends sensitively on both the nature of the inhomogeneities and the velocity of the superfluid. A systematic classification of the dynamical regimes arising under these conditions remains an open problem.

Impurities profoundly shape superfluid dynamics by inducing dissipation akin to classical fluid behavior.
In Bose superfluids, they often serve as nucleation sites for quantum vortices~\cite{PhysRevLett.104.150404,s31t-tjl9}, which are the primary agents coupling the superfluid and normal components and enabling energy transfer between them. For instance, the Hall-Vinen-Bekarevich-Khalatnikov model (a generalization of Landau’s two-fluid theory) requires both a superfluid fraction below unity and the presence of quantized vorticity for dissipation to occur during the dynamics (see~\cite{Barenghi2006-wa} for reviews). Thus, impurities influence superfluid behavior in two fundamental ways: first, their mere presence effectively reduces the superfluid fraction; second, by seeding vortices, they enable the transfer of flow energy into the normal component, which subsequently dissipates it into internal excitations.

Ultracold atomic gases provide a tunable platform for studying superfluid transport. In two-dimensional annular or toroidal traps, persistent currents arise with quantized circulation determined by the winding number $w$~\cite{ryu2007observation,cai2022persistent,pecci2021probing,DelPacePRX2022,Polo2025}. 
In Bose–Einstein condensates (BECs), extensive studies have characterized the stability and dissipation of persistent currents in the presence of impurities or barriers~\cite{PhysRevLett.106.130401,PhysRevLett.104.050404}, where quantized vortices induce phase slips that drive superflow decay. Recent work has shown that introducing controlled Josephson junctions~\cite{Pezz2024} or localized impurities~\cite{kxhani} can paradoxically stabilize these currents by raising the critical decay threshold. This stabilization stems from the multiply-connected nature of toroidal or ring geometry: as the number of weak links increases, the phase drop and the maximum local velocity at each link decrease, thereby suppressing phase slips and enhancing the stability of persistent currents.
In stark contrast, dissipation in fermionic superfluid rings with impurities remains largely unexplored, with prior work limited to single-impurity cases~\cite{Xhani_2025, PhysRevA.109.033306}. BCS fermionic superfluids introduce added complexity through fragile Cooper pairs, susceptible to pair breaking—analogous to depairing currents in superconductors ~\cite{Kunchur2019,macmanus2004,miura2022,Miura2024}. In clean rings (the setup without impurities), this pair-breaking threshold, set by the ring radius and interaction strength \cite{Xhani_2025}, can even trigger vortex nucleation, yielding dynamics qualitatively distinct from those of BECs.
Here, we move beyond the single-impurity paradigm and investigate how impurity density and size control the dissipation of persistent currents in fermionic superfluid rings,  and their role in the competition between pair breaking and vortex mobility. To align with realistic experimental protocols, impurities are introduced dynamically into the system, with their sizes both smaller and larger than the coherence length. 
We show that increasing the impurity density can enhance the critical winding number for vortex emission, as observed in BECs \cite{kxhani}; however, unlike in BECs, this enhancement depends on impurity size and is capped by the clean-ring pair-breaking threshold. 
Below this threshold, the winding number remains constant while impurities accelerate flow decay through enhanced pair breaking that strengthens with their density and size. 
Above the threshold, additional dissipation arises from vortex emission and motion, whose interactions with impurities significantly alter vortex mobility and, consequently, energy loss. Distinct regimes of mobility arise with varying impurity density and size, giving rise to different dissipation mechanisms: at low densities, vortex–impurity scattering promotes radial vortex motion (especially for smaller impurities), accelerating superflow decay; at intermediate densities, pinning effects develop, slowing flow energy dissipation; and at high densities, decay speeds up again potentially accompanied by inter-site vortex hopping for sufficiently large impurities, with the pair breaking becoming the dominant dissipative mechanism.

\vspace{0.2em}
\section{Theoretical model} 
\label{section:theory} 
We study the dynamics of persistent currents in a weakly attractive fermionic superfluid within the Bardeen–Cooper–Schrieffer (BCS) regime. The evolution is described using time-dependent density functional theory for superfluid Fermi gases~\cite{PhysRevA.76.040502,Bulgac2008,Bulgac2012}, in the form of the extended Superfluid Local Density Approximation (SLDAE)~\cite{boulet}. This framework has been successfully applied to Josephson dynamics~\cite{PhysRevLett.130.023003}, persistent currents~\cite{Xhani_2025}, vortex dynamics~\cite{PhysRevLett.130.043001,Bulgac2011,Wlazlowski2018}, Higgs modes~\cite{PhysRevLett.102.085302}, spin-imbalanced systems~\cite{PhysRevLett.97.020402,Bulgac2008,PhysRevA.104.053322}, and quantum turbulence~\cite{PhysRevA.91.031602,PhysRevA.105.013304}.

The time-dependent SLDAE equations are formally equivalent to the time-dependent Bogoliubov–de Gennes (BdG) equations (with $\hbar = m = 1$):
\begin{equation}
\label{eq:bdg-td}
\begin{pmatrix}
    h(\vec{r},t)  
    & \Delta(\vec{r},t) \\
    \Delta^*(\vec{r},t) & 
    -h^*(\vec{r},t) 
\end{pmatrix}
\begin{pmatrix}
    u_n(\vec{r},t) \\
    v_n(\vec{r},t)
\end{pmatrix}
= i \frac{\partial}{\partial t}
\begin{pmatrix}
    u_n(\vec{r},t) \\
    v_n(\vec{r},t)
\end{pmatrix},
\end{equation}
where $(u_n,v_n)^T$ are the Bogoliubov amplitudes.
Unlike the conventional BdG formalism valid only for weak coupling, SLDAE accurately captures the entire BCS–BEC crossover. Here we focus on $a_s k_F=-1$, with $k_F=(3\pi^2\rho_0)^{1/3}$ defined by the bulk density $\rho_0$, and $a_s$ being the scattering length.
The normal, anomalous, and current densities are expressed as
\begin{equation}
    \rho = 2\sum_{n}\lvert v_n\rvert^2, \,\,\,
    \nu  = \sum_{n}u_n v_n^{*}, \,\,\,
    \vec{j}= 2\sum_{n}\textrm{Im}\left(v_n \vec{\nabla} v_n^* \right)
\end{equation}
where the sums run over quasiparticle states with $E_n\in(0,E_c)$, with the cutoff $E_c$ that regularizes ultraviolet divergences~\cite{Bulgac2002,PhysRevA.76.040502}.
The single-particle Hamiltonian and pairing field are
\begin{equation}\label{eq:sldae-current}
  h = -\tfrac12\nabla^{2} + U +V_{\textrm{ext}}(\vec r,t),\qquad
  \Delta = -\frac{C}{\rho^{1/3}}\nu,
\end{equation}
with $U$ the mean-field potential, $V_{\textrm{ext}}$ the external potential (trap and impurity potentials), and $C$ a regularized pairing constant. Their explicit expressions are provided in Ref.~\cite{boulet} and assure proper reproduction of the quantum Monte Carlo equation of state and pairing gap.
For a uniform system at a considered coupling constant, SLDAE gives $\Delta/\varepsilon_F \approx 0.134$, where $\varepsilon_F=k_F^2/2$ is the Fermi energy, and the corresponding coherence length is $\xi=\varepsilon_F/k_F\Delta\approx7.5\,k_F^{-1}$. 

The superfluid is confined in a two-dimensional ring potential with inner and outer radii $R_{\mathrm{in}} = 30 k_F^{-1}$ and $R_{\mathrm{out}} = 60 k_F^{-1}$, respectively. Trapping is applied only in the $x$–$y$ plane, while translational symmetry is imposed along the $z$ direction; thus, the system is effectively three-dimensional at the microscopic level.
A controllable superflow is generated using the phase-imprinting technique. Specifically, a phase $\phi = w_0 \arctan(x, y)$ is imprinted onto the order parameter, $\Delta(x, y) = |\Delta(x, y)| e^{i\phi(x,y)}$, and the corresponding ground state is obtained from stationary simulations [$i\partial/\partial t\rightarrow E_n$ in Eq.~(\ref{eq:bdg-td})] in the absence of impurities. This procedure initializes a rotating superfluid with a total phase winding of $2\pi w_0$, corresponding to a persistent-current state with winding number $w_0$.

A key distinction between BEC and BCS superfluids arises already at this stage: in the latter, it exists an additional threshold, the pair-breaking one defined as $\wpb\approx 2R_{\mathrm{in}}\vpb$, where $\vpb=\sqrt{\sqrt{\mu^2+\Delta^2}-\mu}$ and $\mu$ is the chemical potential, which depends on the inner radius of the ring and interaction strength ~\cite{Xhani_2025}. Above this limit, pair-breaking suppresses the order parameter near the inner radius, potentially leading to vortex emission and the subsequent decay of the current. For the present setup, the estimated threshold $\wpb \approx 5$.

\begin{figure*}[t]
  \centering
\includegraphics[width=0.9\linewidth, trim=0mm 0mm 0mm 0mm, clip]{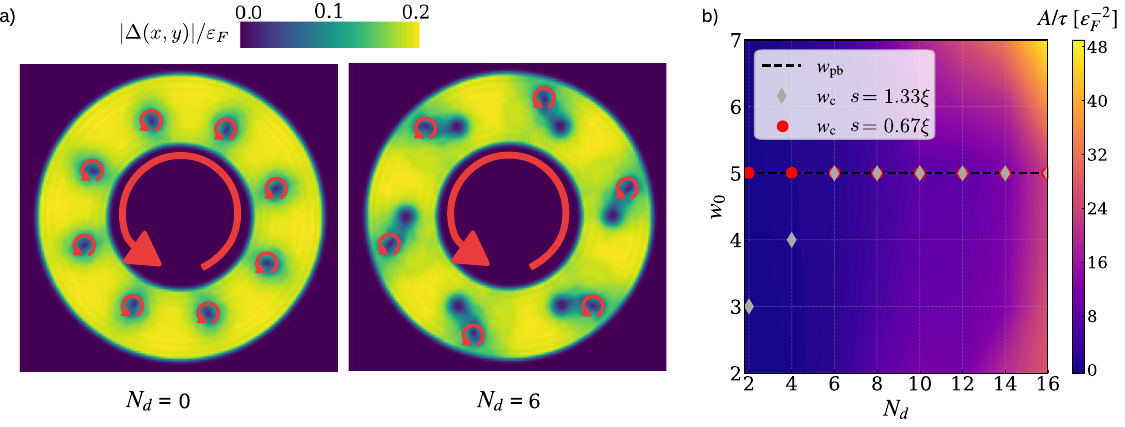}
\caption{(a) Magnitude of the pairing field $|\Delta(x,y)|$ for $w_0 = 7$ in the absence of impurities ($N_d = 0$) and with $N_d = 6$ impurities, shown at $t\varepsilon_F \approx 400$. The impurity size is $s_2=0.67\xi$. The imposed flow direction and vortex positions are indicated by the swirling arrows.
(b) Flow–energy decay rate $A/\tau$ for impurity size $s_1 = 1.33\xi$ as a function of winding number $w_0$ and impurity number $N_d$. Grey diamonds mark the boundary between unstable (winding number decays) and stable ($w(t)$ constant) regimes for $s_1 = 1.33\xi$. Red circles mark the same boundary for $s_2 = 0.67\xi$. The dashed black line indicates the pair-breaking threshold $\wpb$.}
\label{fig:Fig1}
\end{figure*}
Similar to the experimental procedure~\cite{Pezz2024}, impurities are introduced dynamically during the simulation. Each impurity is modeled by a Gaussian potential symmetrically distributed along a ring (forming a circular lattice): 
\begin{align}\label{eq:potential}
V_{\textrm{imp}}(\vec r,t) &=  V_0(t)\sum_{i=1}^{N_d}\exp\!\left[-\frac{(x-x_i)^2+(y-y_i)^2}{2\sigma^2}\right],
\end{align}
where $V_0(t)$ is the time-dependent amplitude, ramped from zero to its final value ($\approx 2\varepsilon_{F}$) within the time interval $(0\text{–}25)\varepsilon_F^{-1}$. 
The impurity centers are positioned at $x_i = R_0\cos(2\pi i/N_d)$ and $y_i = R_0\sin(2\pi i/N_d)$, with $R_0 = (R_{\mathrm{in}} + R_{\mathrm{out}})/2$. We explore the system’s response as a function of two control parameters: the number of impurities $N_d$ and their size $s$, with the latter defined as the full width at half maximum ($s \approx 2.355\sigma$). Two representative regimes are considered:
impurities with size larger ($s_1 = 1.33\xi$) and smaller ($s_2 = 0.67\xi$) than the coherence length. In our simulations, we vary $2 \le N_d \le 16$.

The equations of motion~(\ref{eq:bdg-td}) are solved numerically on a computational grid of size $128\times 128$, with lattice spacing $ d \mathrm{x}k_F = 1$, and plane waves along the third direction are assumed. All simulations are performed using the \verb|W-SLDA Toolkit|~\cite{WSLDAToolkit}.

\vspace{0.2em}
\section{Results} 
\label{section:dynamic}

Representative snapshots from the time-dependent simulations are shown in Fig.~\ref{fig:Fig1}(a) for the case $w_0 = 7$ and for two configurations: without impurities ($N_d = 0$) and with $N_d = 6$ impurities. As $w_0=7$ is larger than the pair-breaking threshold, the persistent current
is unstable even without impurities, as found in \cite{Xhani_2025}. In fact, vortices are spontaneously nucleated (at the inner edge for $N_d = 0$, which next propagate to the bulk), which leads to the decay of the current. In the presence of impurities, instead, vortex motion can be controlled due to vortex-impurity interaction, as is evident in Fig.~1(a) for $N_d=6$. In order to investigate the role of impurity number and sizes on the superflow dissipation mechanisms,
we introduce an energy-based diagnostic by computing the flow energy, which measures the total current in the system:
\begin{equation}
\Eflow(t)=\int \frac{\vec{j}^2(\vec{r},t)}{2\rho(\vec{r},t)}\,d^3 \vec{r}.
\end{equation}
This quantity includes contributions from both the superfluid and normal components, since within the DFT framework $\vec{j} \neq \rho\, (\hbar/2)\nabla\phi$ (the factor $2$ corresponds to the Cooper pair mass). This differs from the Gross–Pitaevskii approach, where the equality holds exactly. Consequently, $\Eflow$ captures both vortex dynamics and normal-fluid dissipation~\cite{Xhani_2025}.
We also monitor the condensation energy,
\begin{equation}
\label{eq:econd}
\Econd(t)= \frac{3}{8} \int  \frac{|\Delta(\vec{r},t)|^2 }{\varepsilon_F(\vec{r},t)}\rho(\vec{r},t)\,d^3\vec{r},
\end{equation}
where $\varepsilon_F(\vec{r})$ is the local Fermi energy~\cite{wlazlowski2023dissipation}. This quantity measures the pairing strength and provides a direct probe of pair-breaking processes during the evolution.
To quantify dissipation, the temporal profile of the flow energy, $\Eflow(t)$, is fitted with an exponential function, $f(t) = A \exp\left[-(t - t_0)/\tau\right]+B$ where $A$ and $B$ are related to initial and final values, $\tau$ the decay time, and $t_0 = 25\,\varepsilon_F^{-1}$ denotes the time when the impurities are fully ramped on. Additionally to that, we evaluate the change in the flow energy per particle, $\delta \Eflow/N = [\Eflow(t_{\textrm{f}}) - \Eflow(t_{\textrm{i}})]/N\eF$ and the relative change in the condensation energy $\delta \Econd = [\Econd(t_{\textrm{f}}) - \Econd(t_{\textrm{i}}) ]/\Econd(t_{\textrm{i}})$, where $t_{\textrm{i}} = 25\varepsilon_F^{-1}$ and $t_{\textrm{f}} = 1000\varepsilon_F^{-1}$.

To characterize the topological stability of the current, we extract the temporal evolution of the average winding number value. We define the critical winding number $w_c$ for vortex emission as the largest initial winding $w_0$ that remains stable over the time interval $1000\,\varepsilon_F^{-1}$.
Fig.~\ref{fig:Fig1}(b) summarizes the dependence of $w_c$ on the number of impurities $N_d$ and on the impurity size.
We first find that the critical winding number $w_c$ increases with $N_d$ for larger impurity sizes, consistent with previous observations in BECs~\cite{kxhani, Pezz2024}. Interestingly, this trend persists only up to the clean-ring pair-breaking threshold, $\wpb=5$. In contrast, for smaller defects ($s_2 = 0.67\xi$), we observe that $w_c$ remains equal to $\wpb$ across the entire range of $N_d$ considered.
This shows that the clean-ring pair-breaking threshold $w_{pb}$ caps the impurity-induced enhancement of the vortex-emission critical winding, with a size-dependent trend. Unlike previous studies on superconductors performed in non-ring geometries with static impurities~\cite{Kunchur2019,macmanus2004,miura2022,Miura2024}—where the critical current grows with impurity density but stays below depairing value~\cite{Kunchur2019,macmanus2004,miura2022,Miura2024}—small impurities here render $w_c$ density-independent, saturating at $w_{pb}$. This behaviour is due to a mixed superfluid–normal state which introduces new physics. Since $w_c$ depends on both the height and width of the impurities \cite{atoms11080109}, we choose these parameters to represent two distinct scenarios, such that $w_c(N_d=2)$ is either equal to or below the pair-breaking threshold. Even in the absence of vortex emission—i.e.~for $w_0 \leq w_c (N_d)$, the colormap in Fig.~1(b) shows that increasing the impurity number produces a finite flow‐energy decay rate, indicating dissipation despite a constant average winding number. To illustrate this behavior, we examine the case $w_0=3$ as a function of $N_d$ [Fig.~2]. For $N_d>2$, the winding number stays constant, yet both the decay rate  $A/\tau$ and the flow‐energy loss $\delta E_{\rm flow}$ increase with $N_d$. This trend stems from impurity–induced pair breaking, which depletes the condensate—as seen from the reduced condensation energy in Fig.~2(d)—and promotes flow‐energy dissipation. The effect strengthens with impurity number and size. Therefore, impurities extend the occurrence of pair-breaking events even below the clean-ring threshold, while still conserving the winding number. 
For $ w_c(N_d=2) \leq  w_0 \leq \wpb$, the system’s topological stability evolves with increasing impurity concentration—from an unstable regime, where the winding number decays in time, to a stable regime with constant w(t). In this range, the superflow remains dissipative, though its microscopic origin depends on $N_d$: for relatively low $N_d$, both vortices and pair‐breaking contribute to flow energy loss, while at higher $N_d$ the dissipation stems exclusively from pair‐breaking, as phase slippage becomes suppressed. 
When $w_0$ exceeds the pair‐breaking threshold (e.g.~, $w_0=7$), topological stability is never achieved, regardless of impurity size or number [Fig.~2(b)]. Increasing the number of impurities accelerates the decay of flow energy and enhances pair‐breaking, with both effects becoming more pronounced for larger impurities. At intermediate impurity concentrations ($6 \leq N_d \leq 12$), dissipation temporarily slows down for both impurity sizes but resumes increasing when $N_d$ grows further, leading again to faster losses of both flow and condensation energies.
Pair‐breaking is always present—as confirmed by finite condensation energy losses that scale with $N_d$ and impurity size. However, these observations alone do not reveal details about vortex dynamics. To clarify this aspect, we extracted the vortex trajectories and mobility as a function of impurity number and size, finding that vortex-impurity interactions create distinct mobility regimes that govern dissipation.

\begin{figure}[t]
  \centering
    \includegraphics[width=\columnwidth, trim=0mm 0mm 0mm 0mm, clip]{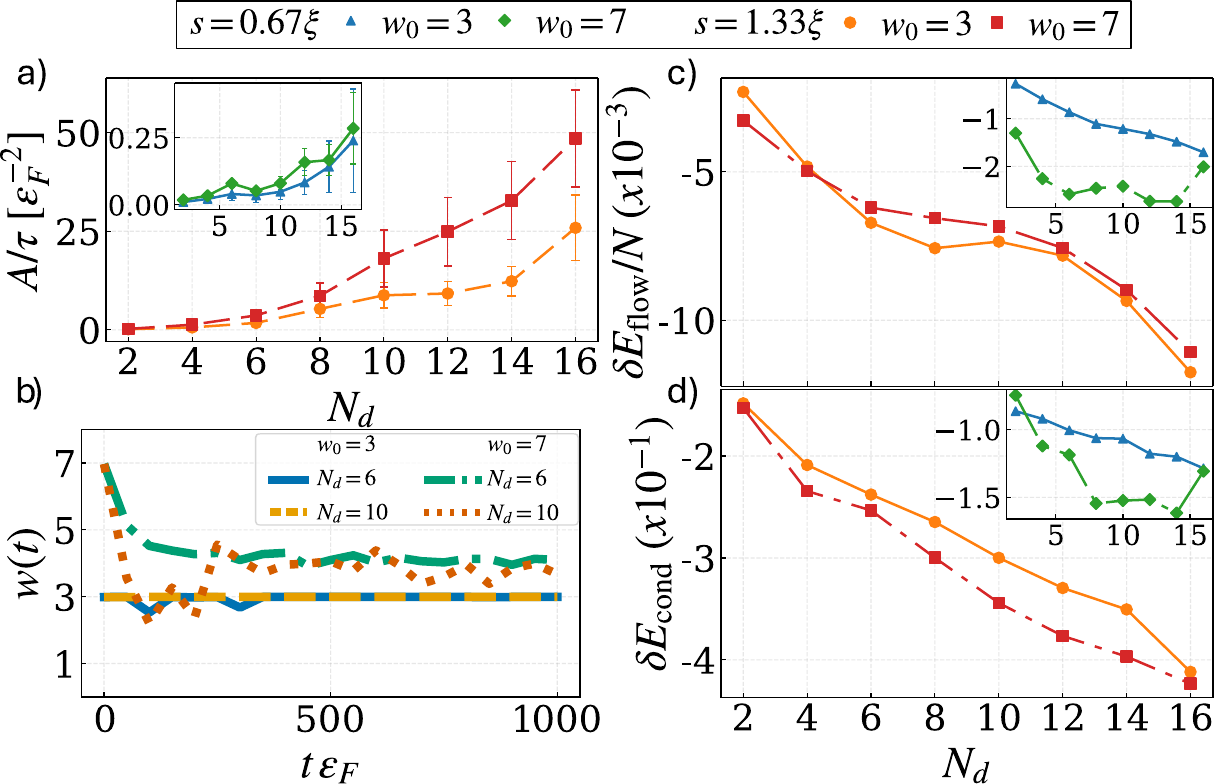}
\caption{Dissipation mechanisms of persistent currents in the presence of impurities.
(a) Flow–energy decay rate $A/\tau$ obtained from exponential fits for impurity sizes $s = 1.33\xi$ (main panel) and $s = 0.67\xi$ (inset).
(b) Time evolution of the winding number $w(t)$ for $s = 1.33\xi$.
(c) Per-particle flow–energy change $\delta \Eflow/N$ between initial and final states.
(d) Condensation–energy change $\delta \Econd$ highlighting pair-breaking effects. Panels (c) and (d) show that the impurity size primarily governs variations in the flow and condensation energies, rather than the topological properties of the order parameter, which in turn are encoded in $w(t)$. }
\label{fig:Fig2}
\end{figure}

\begin{figure}[t]
\includegraphics[width=2\columnwidth, trim=0mm 176mm 0mm 0mm,]{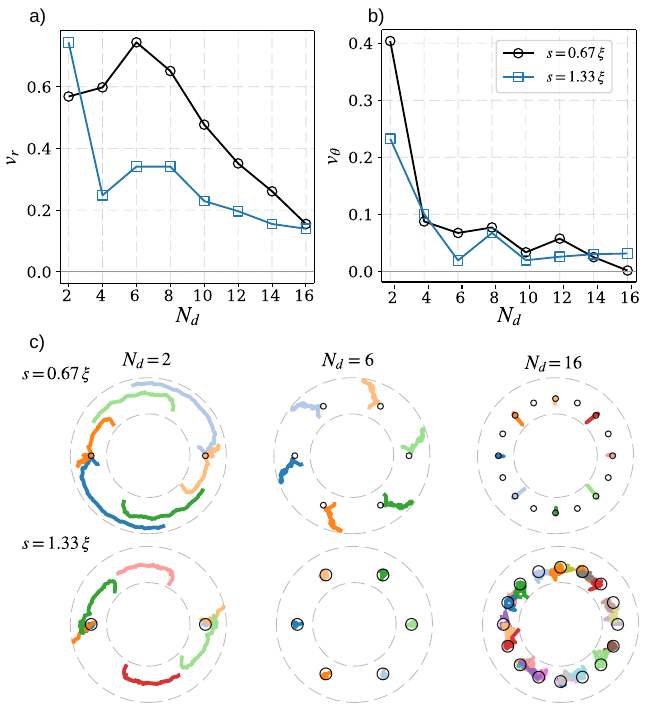}
\caption{(a,b) Radial and angular vortex mobilities as functions of the number of impurities $N_d$ for two impurity sizes, at $w_0 = 7$.
(c) Two-dimensional vortex trajectories for $w_0 = 7$ for different numbers of impurities and their sizes. Different colors indicate different vortices.}
\label{fig:Fig3}
\end{figure}
 
We calculate the radial and angular mobilities, shown in Fig.~\ref{fig:Fig3}(a,b) for a fixed initial winding number $w_0 = 7$, defined as
\begin{align}
v_r &= \frac{\overline{r}}{N_v}\sum_{k=1}^{N_v}\frac{\Delta r_k}{\Delta t_k}, \quad
v_\theta = \frac{\overline{r}^2}{N_v}\sum_{k=1}^{N_v}\frac{\Delta\theta_k}{\Delta t_k},
\end{align}
where $N_v$ is the number of vortices, $\overline{r} = \frac{1}{N_v}\sum_{k} r_{k}^{(\textrm{i})}$ is the mean initial radius, and $\Delta r_k = r_{k}^{(\textrm{f})}-r_{k}^{(\textrm{i})}$, $\Delta \theta_k = \theta_{k}^{(\textrm{f})}-\theta_{k}^{(\textrm{i})}$, $\Delta t_k = t_{k}^{(\textrm{f})}-t_{k}^{(\textrm{i})}$ denote the radial, angular, and temporal displacements between the initial (i) moment of vortex nucleation and its annihilation or the end of the simulation (f). 

Impurities strongly influence both vortex emission and motion, with trajectories determined by defect size and spacing. In clean systems, vortices are symmetrically emitted along the inner edge of the ring, forming a periodic, freely propagating pattern [Fig.~1(a) left panel]. The introduction of impurities breaks this symmetry: for $N_d=2$, weak scatterers deflect vortex trajectories, whereas stronger ones induce pinning of some of the vortices and trigger the emission of additional vortices, as visible in Fig.~3(c). As impurity density increases, the radial mobility exhibits opposite trends for the two impurity sizes. For larger defects, both radial and angular mobilities decrease steadily with $N_d$, as strong pinning centers restrict the motion of all emitted vortices (see $N_d=6$ in Fig.~3(c)). For smaller impurities, however, the radial mobility $v_r$ initially rises up to $N_d=6$: vortex–impurity scattering deflects trajectories outward, forcing vortices to traverse a broader superfluid region and thus enhancing flow dissipation.
Contrary to standard expectations, even pinned vortices exhibit finite flow dissipation, as shown in Fig.~2(c). This effect stems from impurity-induced pair-breaking processes. Interestingly, at moderate impurity densities ($N_d>6$), dissipation slows down, particularly for smaller defects—manifested as a plateau in $\delta \Eflow$ for $s_2 = 0.67\xi$ and $w_0 = 7$—signaling a partial stabilization of flow energy with increasing $N_d$. At higher densities ($N_d=16$) and for larger impurities, vortices begin tunneling between defect sites, with the angular mobility $v_{\theta}$ saturating at a finite value. This behavior indicates a qualitatively distinct dynamical regime in which vortices remain locally pinned but acquire collective mobility through inter-site hopping.
These observations reveal four distinct mobility regimes—deflected trajectories, individual pinning, collective pinning, and inter-site hopping—absent in clean systems. They represent emergent mechanisms governing vortex dynamics and dissipation in fermionic superfluids.

\vspace{0.2em}

\section{Conclusions}
In conclusion, we demonstrate that impurity size and density precisely control the balance between pair breaking and vortex emission/dynamics—the two primary dissipation channels for persistent currents in BCS-regime fermionic superfluid rings. Time-dependent density functional theory reveals how these parameters tune both mechanisms, the vortex-emission critical winding $w_c$ and vortex mobility. Below $w_c$, denser or larger impurities enhance pair breaking, causing flow energy decay without winding number loss, unlike clean rings. Above $w_c$, the vortex-impurity interaction, defined by the interplay of impurity spacing, size, and coherence length, governs whether vortices are deflected, individually or collectively pinned, or tunnel between impurity sites. These dynamics shift the dissipation mechanism—from pair breaking plus mobile vortices at low $N_d$ to dominant pair breaking at higher densities.
Impurity size further sets the dependence of $w_c$ on $N_d$: small impurities yield density-independent critical winding, while large ones progressively enhance it toward the clean-ring pair-breaking limit. Impurity engineering thus provides targeted control of superfluid transport, with implications for ultracold atoms and neutron-star superfluids.

Although neutron matter is strongly interacting, its pairing gap and coherence length correspond to a weak-coupling BCS regime~\cite{Carlson2012}. The impurity spacings explored here are comparable to the expected distances between nuclear clusters in the neutron-star crust~\cite{Negele1973,PhysRevC.85.035801}. Hence, the mechanisms identified (pair-breaking dissipation and impurity-mediated pinning) offer insight into vortex dynamics in neutron-star superfluids and their role in phenomena such as pulsar glitches~\cite{Chamel2008-so,Haskell2015-ct,Antonopoulou2022}.

\section{Acknowledgements}
We thank Brynmor Haskell for fruitful discussions.
This work was financially supported by the (Polish) National Science Center Grants No. 2022/45/B/ST2/00358.
K. X. acknowledges funding from the Italian MUR (PRIN DiQut Grant No. 2022523NA7).
We acknowledge the Polish high-performance computing infrastructure, PLGrid, for awarding this project access to the LUMI supercomputer, owned by the EuroHPC Joint Undertaking, hosted by CSC (Finland), and the LUMI consortium through PLL/2024/07/017603.
\paragraph*{Author contributions:}
The work was conceived by K.X. Calculations were executed by A.B., and data analysis was performed by B.T. All authors contributed to research planning, interpretation of the results, and manuscript writing. 
\paragraph*{Data avalibility}
The data that support the findings of this article are openly available~\cite{zenodo}.
\bibliography{FJJ}
\end{document}